\documentstyle[aps]{revtex}

\begin{document}

\title{The ground state of the carbon atom in strong
magnetic fields}
\author{M. V. Ivanov\dag\  and P. Schmelcher}
\address{Theoretische Chemie, Physikalisch--Chemisches Institut,
Universit\"at Heidelberg, INF 253, D-69120 Heidelberg,
Federal Republic of Germany\\
\dag Permanent address: Institute of Precambrian Geology and Geochronology,
Russian Academy of Sciences,
Nab. Makarova 2, St. Petersburg 199034, Russia
}

\date{\today}
\maketitle

\begin{abstract}
The ground and a few excited states of the carbon atom
in external uniform magnetic fields are calculated
by means of our 2D mesh Hartree-Fock method for field strengths
ranging from zero up to $2.35\cdot 10^9$T.
With increasing field strength the ground state undergoes six 
transitions involving seven different electronic configurations 
which belong to three groups with different spin projections 
$S_z=-1,-2,-3$. For weak fields the ground state configuration arises from the
field-free $1s^2 2s^2 2p_0 2p_{-1}$, $S_z=-1$ configuration.
With increasing field strength the ground state involves the 
four $S_z=-2$ configurations  
$1s^22s2p_0 2p_{-1}2p_{+1}$, 
$1s^22s2p_0 2p_{-1}3d_{-2}$, 
$1s^22p_0 2p_{-1}3d_{-2}4f_{-3}$ and 
$1s^22p_{-1}3d_{-2}4f_{-3}5g_{-4}$, 
followed by the two fully spin polarized $S_z=-3$ configurations  
$1s2p_02p_{-1}3d_{-2}4f_{-3}5g_{-4}$ and 
$1s2p_{-1}3d_{-2}4f_{-3}5g_{-4}6h_{-5}$. 
The last configuration forms the ground state of the carbon atom in the high field regime 
$\gamma>18.664$. The above series of ground state configurations is extracted from the results of 
numerical calculations for more than twenty electronic configurations selected due to
some general energetical arguments.
\end{abstract}


\section{Introduction}
 
The behaviour and properties of atoms in strong magnetic fields
is a subject of increasing interest. On the one hand this is
motivated by the astrophysical discovery of strong fields
on white dwarfs and neutron stars \cite{NStar1,NStar2,Whdwarf1}
and on the other hand the competition of the diamagnetic and Coulombic
interaction causes a rich variety of complex properties which are
of interest on their own.
The carbon atom, which is the subject of the present investigation, plays a 
major role for the evolution of stars and is also expected to occur
in the case of magnetic white dwarfs and neutron stars. 

Investigations on the electronic structure in the presence of
a magnetic field appear to be quite complicated
due to the intricate geometry of this quantum problem. 
Most of the investigations in the literature 
focused on the hydrogen atom 
(for a list of references see, for example, \cite{RWHR,Ivanov88,Fri89,Kra96}).
The results of these studies provided us with an understanding
of the absorption features of certain magnetic white dwarfs and
allowed for a modeling of their atmospheres (see ref.\cite{Rud94} for a 
comprehensive review up to 1994 and ref.\cite{Schm98} for a more recent review on
atoms and molecules in strong magnetic fields).
On the other hand there is a number of magnetic white dwarfs whose
spectra remain unexplained and cannot be interpreted in terms of
magnetized atomic hydrogen. Furthermore new magnetic objects 
are discovered (see, for example, Reimers et al \cite{Reim98} 
in the course of the Hamburg ESO survey) whose spectra await to be explained.  
The most prominent of the unexplained magnetic objects is the white dwarf GD 229.
Very recently significant progress has been achieved with respect to the
interpretation of its rich spectrum ranging from the UV to the near IR.
Extensive and precise calculations on the helium atom provided data
for many excited states in a broad range of field strengths \cite{Bec99}. The comparison
of the stationary transitions of the atom with the positions of the absorption edges
of the observed spectrum yielded strong evidence for the existence of helium
in the atmosphere of GD229 \cite{JSBS229}. 

For the hydrogen atom the impact of the competing Coulomb and diamagnetic interaction
is particularly evident and pronounced in the intermediate regime for which the magnetic
and Coulomb forces are comparable.  
For different electronic degrees of excitation of the atom the intermediate
regime is met for different absolute values of the field strength.
For the ground state this regime is roughly given by $\gamma=~0.2-20$  ($\gamma =B/B_0$
is the magnetic field strength in atomic units, $B_0=\hbar c/ea_0^2=2.3505 {\cdot} 10^5$T).
Both early \cite{Garstang} and more recent works
\cite{RWHR,SimVir78,Friedrich,Fonte,Schmidt}
on the hydrogen atom have used different approaches for relatively
weak fields (the Coulomb force prevails over the magnetic force)
and for very strong fields (the Coulomb force can be
considered as weak in comparison with the magnetic forces which is the so-called
adiabatic regime). In the latter regime the motion of the electron parallel to the magnetic field is
dominated \cite{ElLoudon} by a 1D quasi-Coulomb
potential including a parameter which depends on the magnetic field strength.
The detailed calculations of the hydrogen energy levels carried out
by R\"osner {\it et al} \cite{RWHR} also retained the separation into
different regimes of the field strength by decomposing
the electronic wave function either in terms of spherical
(weak to intermediate fields) or cylindrical (intermediate to high fields) orbitals.
A powerful method to obtain comprehensive results on low-lying energy levels of the
hydrogen atom in particular in the intermediate regime is provided by
mesh methods \cite{Ivanov88}.

For atoms with several electrons there are two decisive factors which 
enrich the possible changes in the electronic structure with varying
field strength compared to the one-electron system. 
First we have a third competing interaction which is
the electron-electron repulsion and second the different electrons
feel very different Coulomb forces, i.e. possess different one particle energies,
and consequently the regime of the intermediate field strengths
appears to be the sum of the intermediate regimes for the separate electrons.

There exist a number of investigations on two-electron atoms in the literature
\cite{Mueller75,Virtamo76,Larsen,Gadiyak,Neuhauser,VinBay89,Ivanov91,TKBHRW,Ivanov94,JonesOrtiz,JonesOrtiz97}.
The majority of them deals with the adiabatic regime in superstrong fields and
the early works are mostly Hartree-Fock (HF) type calculations.
There are also several early variational calculations for the low-field domain
\cite{Larsen,Henry74,Surmelian74}. HF calculations for arbitrary field strengths
have been carried out in refs.\cite{RWHR,TKBHRW} by applying
two different sets of basis functions in the high- and low-field regimes. As a result
of the complicated geometry this approach suffers in the intermediate regime 
from very slow convergence and low accuracy of the calculated energy eigenvalues.
Accurate HF calculations for arbitrary field strengths were carried out in refs.
\cite{Ivanov91,Ivanov94} by the 2D mesh HF method. Investigations on the
ground state as well as a number of excited states of helium including
the correlation energy have recently been performed via a Quantum Monte
Carlo approach \cite{JonesOrtiz97}. Very recently benchmark results with
a precision of $10^{-4}-10^{-6}$ for the energy levels
have been obtained for a large number of
excited states with different symmetries using a configuration
interaction approach with an anisotropic Gaussian basis set \cite{Bec99}.

Focusing on systems with more than two electrons however the number of investigations is very 
scarce \cite{Neuhauser,JonesOrtiz,Muell84,Ivanov98,IvaSchm98}.
In view of the above there is a need for further quantum mechanical
investigations and for data on atoms with more than two electrons in a strong magnetic field.
For the carbon atom there exist only two investigations \cite{Neuhauser} in the adiabatic approximation
 which give a few values
for the binding energies in the high field regime and one more relevant recent
work by Jones {\it et al} \cite{JonesOrtiz}.
The latter contains Hartree-Fock calculations for 
three states of the carbon atom in magnetic fields 
from $\gamma=0.0072$ up to $\gamma=21.6$.
The analysis of these results and in particular their comparison 
with our results are presented in sections IV and V. 

In the current work we apply a fully numerical 2D Hartree-Fock method
to the problem of the carbon atom in magnetic fields and obtain for the first
time conclusive results on the ground state configurations for arbitrary
field strengths. Our approach enables us to perform calculations for various states with
approximately equal precision for weak, intermediate and
high magnetic fields. To identify the ground state for arbitrary field strengths
both general considerations and electronic structure calculations have to be
performed.

\section{Computational Method}

We solve the electronic Schr\"odinger equation for the carbon atom in
a magnetic field under the assumption of an infinitely heavy nucleus
in the (unrestricted) Hartree-Fock approximation.
The numerical approach applied in the present work coincides with that
of our previous investigations \cite{Ivanov98,IvaSchm98}. The latter contain
some more details of the mesh techniques.
The solution is established in the cylindrical coordinate system
$(\rho,\phi,z)$ with the $z$-axis oriented along the magnetic field.
We prescribe to each electron a definite value of the magnetic
quantum number $m_\mu$.
Each one electron wave function $\Psi_\mu$ depends on the variables
$\phi$ and $(\rho,z)$ as follows
\begin{eqnarray}
\Psi_\mu(\rho,\phi,z)=(2\pi)^{-1/2}e^{-i m_\mu\phi}\psi_\mu(z,\rho)
\label{eq:phiout}
\end{eqnarray}
where $\mu$ indicates the numbering of the electrons.
The resulting partial differential equations for $\psi_\mu(z,\rho)$
and the formulae for the Coulomb and exchange potentials
have been presented in
ref.\cite{Ivanov94}.

The one-particle equations for the wave functions $\psi_\mu(z,\rho)$
are solved
by means of the fully numerical mesh method described in refs.
\cite{Ivanov88,Ivanov94}.
The feature which distinguishes the present calculations from
those described in ref.\cite{Ivanov94} is the method of calculation
of the Coulomb and exchange integrals. In the present work as well as in
ref.\cite{Ivanov98,IvaSchm98}
we obtain these potentials as solutions
of the corresponding Poisson equation.

Our mesh approach is flexible enough to yield precise
results for arbitrary field strengths.
Some minor decrease of the precision
appears in very strong magnetic fields.
This phenomenon
is due to a growing difference in the
binding energies ${\epsilon_B}_\mu$
of one electron wave functions belonging to the same
electronic configuration
\begin{eqnarray}
{\epsilon_B}_\mu=(m_\mu+|m_\mu|+2s_{z\mu}+1)\gamma/2-\epsilon_\mu
\label{eq:ebinone}
\end{eqnarray}
where $\epsilon_\mu$ is the one electron energy
and $s_{z\mu}$ is the spin $z$-projection.
The precision of our results depends, of course, on the number of the mesh nodes
and can be improved in calculations with denser meshes.
Most of the present calculations 
are carried out on sequences of meshes with the maximal number of nodes being
$65 \times 65$. 

It was demonstrated in ref.\cite{Ivanov98} that the one electron wave functions 
obtained in multi-electron Hartree-Fock mesh calculations 
can for some atomic states possess a lower spatial symmetry 
than the symmetry constrained traditional HF approaches based on
basis sets. For example it was shown \cite{Ivanov98}
for the $1s^2 2s^2$ state of the beryllium atom 
that the wave functions of the $2s^2$ 
electrons reveal a broken spatial symmetry with respect to the $z=0$ plane. 
The contribution of this effect to the total energy
was found to be significant
for $\gamma>0.5$. In the case of the carbon atom the $2s^2$ electron pair 
belongs to the ground state configuration for the regime $\gamma\leq 0.1862$ (see table \ref{tab:grcon})
for which the symmetry breaking effect does not occur.
On the other hand, we do not expect such kind of a broken symmetry 
for the ground state configurations at strong magnetic fields because
they involve wave functions with large absolute values of the magnetic
quantum numbers. Due to these reasons the present calculations
are based on one electron wave functions with a definite $z$-parity $\pi_z=\pm 1$.

\section{Ground state electronic configurations for 
$\gamma=0$ and $\gamma\rightarrow\infty$}

Let us provide some helpful qualitative
considerations on the problem of the atomic multi-electron ground states
in the limit of strong magnetic fields.
It is clear that the field-free ground state of the carbon atom remains the
ground state only for relatively weak fields.
The set of one electron wave functions constituting
the HF ground state for the opposite case of extremely strong
magnetic fields can be determined as follows.
The nuclear attraction energies and HF potentials
(which determine the motion along $z$ axis)
are for large $\gamma$ small compared to the interaction energies with the magnetic field
(which determines the motion perpendicular to the magnetic field
and is responsible for the Landau zonal structure of the spectrum).
Thus, all the one electron wave functions must correspond to the
lowest Landau zones, i.e. $m_\mu\leq 0$
for all the electrons, and the system must be fully spin-polarized,
i.e. $s_{z\mu}= -{1\over2}$.
For the Coulomb central field the one electron levels form
quasi 1D Coulomb series with the binding energy
$E_B={1\over{2n_z^2}}$ for $n_z>0$ and
$E_B\rightarrow \infty$ for $n_z=0$,
where $n_z$ is the number of nodal
surfaces of the wave function crossing the $z$ axis.

Considering the case $\gamma \rightarrow \infty$ it is evident, that the wave functions
with $n_z=0$ have to be chosen for the ground state configuration.
Furthermore starting with the energetically lowest one particle
level the electrons occupy according to the above arguments orbitals with increasing
absolute value of the magnetic quantum number $m_{\mu}$.
Consequently the ground state of the carbon atom must be given by the configuration
$1s2p_{-1}3d_{-2}4f_{-3}5g_{-4}6h_{-5}$.

For the opposite case $\gamma=0$ it is well-known
that the ground state of the carbon atom corresponds to the spectroscopic term $^3P$. 
In the framework of the non-relativistic consideration 
this term consists of nine states 
degenerate due to three possible $z$-projections of the total spin $S_z=-1,0,1$ 
and three possible values of the total magnetic quantum number $M=-1,0,1$. 
Conventional Hartree-Fock calculations provide the following values 
for the energy for this term: $E=-37.688612$ (Clementi and Roetti in \cite{Clementi}) 
and $E=-37.68880$ (Froese Fischer in \cite{FFischer72}). 

The problem of the configurations of the
ground state for the intermediate field region cannot be
solved without doing explicit calculations 
combined with some qualitative considerations in order to
extract the relevant configurations.
With respect to our notation of the configurations we implicitly assume in the
following that all paired electrons, like
for example the $1s^2$ part of a configuration, are of course in a 
spin up and spin down orbital, respectively, whereas all
unpaired electrons possess a negative projection of the
spin onto the magnetic field direction.

\section{Ground state electronic configurations 
for arbitrary field strengths}

First of all, we divide the possible ground state configurations into three groups  
according to their total spin projection $S_z$ :
the $S_z=-1$ group (low-field ground state configurations), 
the intermediate group $S_z=-2$ and the 
$S_z=-3$ group (the high-field ground state configurations). 
This grouping is required for the following qualitative considerations which are based on the geometry 
of the spatial parts of the one electron wave functions. 

To begin we would like to remark on our calculations for 
the atom without field compared to the traditional Hartree-Fock calculations. 
The Hartree-Fock energies for the $^3P$ 
ground state of the field-free carbon atom have been given above. 
One important feature of the conventional Hartree-Fock calculations \cite{Clementi,FFischer72}
is the correspondence of each one electron wave function 
to only one spherical harmonic. 
As shown for example in ref.\cite{Ivanov98} this restriction does not allow obtaining 
energies which correspond to the Hartree-Fock limit in the sense of a 
fully free variation of the one-particle functions respecting
the exact symmetries of the total system. Not imposing the symmetries 
of the spherical harmonics on the one-particle functions
provides lower Hartree-Fock energies. 
This can be done, for instance, in 2D or 3D mesh Hartree-Fock calculations.  
In the framework of our 2D mesh calculations 
the components of the multiplet $^3P$ are built up 
of different one particle wave functions possessing
as good quantum numbers the magnetic quantum number $m_\nu$,
the $z$-parity and the spin projection $s_z$.  
Due to the higher flexibility of the one particle wave functions 
our Hartree-Fock energies for all the components of 
the $^3P$ multiplet are lower in energy and are slightly different for different 
components of the multiplet. 
At $\gamma=0$ the value of $S_z$ does not affect the energy, 
whereas the spatial parts can contain three valence configurations: 
$2p_0 2p_{-1}$, $2p_0 2p_{+1}$ and $2p_{-1} 2p_{+1}$ 
(the core part $1s^2 2s^2$ of the total configuration is omitted). 
Our energies for the $2p_0 2p_{-1}$ and $2p_0 2p_{+1}$ configurations 
coincide and are $E=-37.69096$, 
whereas the calculation for $2p_{-1} 2p_{+1}$ gives a slightly 
different value namely $E=-37.69376$.  
It is evident that the energies of these configurations must coincide in 3D
HF calculations. On the other hand only the $1s^2 2s^2 2p_0 2p_{-1}$ configuration represents the
energetically lowest component of the $^3P$ multiplet in weak
magnetic fields. We therefore neglect here and in the following the small difference in energy of the
$1s^2 2s^2 2p_0 2p_{+1}$, $1s^2 2s^2 2p_{-1} 2p_{+1}$ 
$1s^2 2s^2 2p_0 2p_{-1}$ configurations in the absence of a 
magnetic field.

According to the arguments presented in the previous section 
we know that the ground state configuration of the carbon atom
in the high field limit must be the fully spin-polarized state  
$1s2p_{-1}3d_{-2}4f_{-3}5g_{-4}6h_{-5}$.
The question of the ground state configurations at intermediate fields 
cannot be solved without performing explicit electronic structure calculations. 
On the other hand, the a priori set of possible intermediate ground 
state configurations increases enormously with increasing number of electrons
and is already for the carbon atom too many in order to perform explicit
calculations for all of them. Some qualitative considerations are therefore
needed in order to exclude certain configurations as possible ground
state configurations thereby reducing the number of candidates for which
explicit calculations have to be performed.  
The optimal strategy hereby consists of the repeated procedure
of determining neighbouring ground state configurations with
increasing (or decreasing) magnetic field strength using both qualitative
arguments (see below) as well as the results
of the calculations for concrete configurations.

The energies for all the considered states and in particular those
of the ground states are illustrated in figures 1-4. 
Figure 1 shows the total energies for the considered configurations
with $S_z=-1$. Figure 2 and 3 illustrate the intermediate ground
state scenario for the relevant configurations possessing the spin projection
$S_z=-2$. Figure 4 shows the relevant fully spin polarized configurations,
i.e. the states with $S_z=-3$. The reader should note
that for each group of configurations the figures illustrate a different
regime of field strengths and energies.
Finally we present in figure 5 the global scenario of the total energies of the ground states 
for the range of considered field strength $0<\gamma<100a.u.$ 
The vertical lines shown in this picture divide the complete field strength regime into
different regions for which different electronic configurations represent the ground state. 
The numerical values for the transition field strengths and the total
energies at which the crossover between different electronic configurations takes place
are given in table \ref{tab:grcon}. In the following paragraphs we describe in detail 
our selection procedure for the candidates of the electronic ground state configurations.

We start our consideration for $\gamma\ne0$  with the high-field ground state and 
subsequently consider other possible candidates in question for the electronic ground state for $S_z=-3$ 
(see figure 4) {\it{with decreasing field strength}}.
All the one electron wave functions of the high-field ground state 
$1s2p_{-1}3d_{-2}4f_{-3}5g_{-4}6h_{-5}$
possess no nodal surfaces crossing the $z$-axis and occupy the energetically lowest orbitals
with magnetic quantum numbers ranging from $m=0$ down to $m=-5$. 
We shall refer to the number of the nodal surfaces crossing the $z$ axis 
as $n_z$. The $6h_{-5}$ orbital possesses the smallest binding energy of all orbitals constituting
the high-field ground state. Its binding energy decreases rapidly with decreasing
field strength.  Thus, we can expect that the first crossover of ground state configurations
happens due to a change of the $6h_{-5}$ orbital into one
possessing a higher binding energy at the corresponding lowered range of field strength. 
It is natural to suppose that the first transition while decreasing the magnetic field strength
will involve a transition from an orbital possessing $n_z=0$ to one for $n_z=1$.
The energetically lowest available one particle state with $n_z=1$ is the $2p_0$ orbital. 
Another possible orbital into which the $6h_{-5}$ wave function could evolve is the $2s$ state. 
For the hydrogen atom or hydrogen-like ions in a magnetic field the $2p_{0}$ is stronger bound than the
$2s$ orbital. On the other hand, owing to the electron screening in multielectron atoms in field-free space
the $2s$ orbital tends to be more tightly bound than the $2p_0$ orbital.
Thus, two states i.e. the $1s2p_02p_{-1}3d_{-2}4f_{-3}5g_{-4}$ state as well as the
$1s2s2p_{-1}3d_{-2}4f_{-3}5g_{-4}$ configuration are candidates for becoming the ground state
in the $S_z=-3$ set when we lower the field strength coming from the high field situation.

Analogous arguments lead to the three following 
candidates for the ground state in case of the second crossover in the $S_z=-3$ subset which 
takes place with decreasing field strength:
$1s2s2p_0 2p_{-1}3d_{-2}4f_{-3}$, $1s2p_0 2p_{-1}3d_{-1}3d_{-2}4f_{-3}$ and 
$1s2s2p_{-1}3d_{-1}3d_{-2}4f_{-3}$. 
It is evident that the one particle energies for the $3d_{-1}$ and $2p_0$ obey
$E_{3d_{-1}} > E_{2p_0}$ for all values of $\gamma$ since they possess the same nodal structure
with respect to the z-axis and only the $3d_{-1}$ possesses an additional node in the plane
perpendicular to the z-axis. For this reason the configuration
$1s2s2p_{-1}3d_{-1}3d_{-2}4f_{-3}$ can be excluded from our considerations of the ground state.
This conclusion is fully confirmed by our calculations. 

The final picture of the ground state crossovers in the subset of configurations with $S_z=-3$ 
obtained within our HF calculations is the following: 
At $\gamma \approx 18.664$ the total energy $E(\gamma)$ for the
$1s2p_0 2p_{-1}3d_{-2}4f_{-3}5g_{-4}$ configuration
intersects the total energy curve for the high-field ground state $1s2p_{-1}3d_{-2}4f_{-3}5g_{-4}6h_{-5}$ 
and for $\gamma < 18.664$ the energy of the previous state is therefore
lower than the energy of the high-field ground state. 
The $1s2p_0 2p_{-1}3d_{-2}4f_{-3}5g_{-4}$ configuration
remains the energetically lowest one among all the six above-considered 
states with $S_z=-3$ in the regime $6.8<\gamma<18.664$. For $\gamma < 6.8$ the configuration
$1s2s2p_0 2p_{-1}3d_{-2}4f_{-3}$ becomes the lowest one within the $S_z=-3$ subset. 
However, this second crossover within the subset of states with $S_z=-3$
takes place in a regime of field strengths for which the global ground
state configuration of the carbon atom belongs already to the subset $S_z=-2$. 
(See the state $1s^22p_{-1}3d_{-2}4f_{-3}5g_{-4}$ in figure 4.) 
It should be noted that the structure of the wave functions with $S_z=-3$ is simpler 
than in the two other spin subsets which we consider in the following. 
 
The considerations for the $S_z=-2$ subset of states whose energies are illustrated in figures 2 and 3,
are similar to the ones of the subset $S_z=-3$. The most tightly bound state in the limit of high fields is
given by the $1s^22p_{-1}3d_{-2}4f_{-3}5g_{-4}$ configuration.
When decreasing the field strength this configuration can be replaced by
$1s^22p_0 2p_{-1}3d_{-2}4f_{-3}$ or by $1s^22s2p_{-1}3d_{-2}4f_{-3}$ (see figure 3).
The next change could lead to the configurations $1s^22s2p_0 2p_{-1}3d_{-2}$,
$1s^22p_0 2p_{-1}3d_{-1}3d_{-2}$ or $1s^22s2p_{-1}3d_{-1}3d_{-2}$ (figure 3).
In analogy to our argumentation with the subset of states with $S_z=-3$ it is now obvious that
$ E_{1s^22s2p_0 2p_{-1}3d_{-2}} < E_{1s^22s2p_{-1}3d_{-1}3d_{-2}} $
and the latter state cannot become the ground state configuration.
Our electronic structure calculations provide the following sequence of ground states for decreasing field
strength: for high fields we have the $1s^22p_{-1}3d_{-2}4f_{-3}5g_{-4}$ configuration,
with decreasing field strength
a crossover to the configuration $1s^22p_0 2p_{-1}3d_{-2}4f_{-3}$ and with further decreasing
field strength the configuration $1s^22s2p_0 2p_{-1}3d_{-2}$.
Having in mind these numerical results which are presented in figure 3 we provide next an analysis
for further decreasing field strength. It is clear that the orbitals $1s^2$ and $2s2p_0 2p_{-1}$ will
be retained for further decreasing field strength in the framework of the $S_z=-2$ subspace.
Thus the following transitions can affect only the orbital $3d_{-2}$.
This orbital could be changed to the orbitals $3d_{-1}$, $3d_0$, $3p_{-1}$,
$3s$ or $2p_{+1}$.  The $d$-orbitals of this series can be excluded from our consideration,
since they have no preferences over $3d_{-2}$ with respect to the energy neither in
strong nor in weak fields and therefore calculations have to be performed only for the states
$1s^22s2p_0 2p_{-1}3p_{-1}$, $1s^22s2p_0 2p_{-1}3s$ and $1s^22s2p_0 2p_{-1}2p_{+1}$.
Explicit calculations (see figure 2) show that the ground state is given by the $1s^22s2p_0 2p_{-1}2p_{+1}$
configuration.  It is evident that the latter configuration is the ground
state configuration of the subset $S_z=-2$ for the weak field regime and in particular
for $\gamma=0$ and thus it is the last in the sequence of the ground state configurations
with $S_z=-2$. As it turns out in our HF calculations this state as well as all
other states which are ground states for $S_z=-2$ turn out to be
the ground state of the carbon atom as a whole, i.e. taking into account all spin manifolds
$S_z$, for certain regimes of the field strength.

It is reasonable to start the considerations for the configurations with $S_z=-1$
which includes the weak field ground state configuration
$1s^2 2s^2 2p_0 2p_{-1}$ by gradually increasing the field strength
from $\gamma=0$. The configuration $1s^2 2s^2 2p_0 2p_{-1}$ is rather ´robust´ and the most
reasonable transition which could occur in the framework of the configurations with
$S_z=-1$ is the transition of one of the $2s$ electrons to the $2p_{-1}$ orbital i.e. the transition to 
configuration $1s^2 2s2p_0 2p_{-1}^2$. 
The argument therefore is that at $\gamma=0$ the $2s$ orbital 
is more bound than the $2p_0$ and $2p_{-1}$ and for increasing the field the binding energy 
of the $2p_{-1}$ orbital increases very rapidly.  Our explicit calculations (see figure 1) show that 
this transition takes place at $\gamma \approx 0.7$, i.e.  in the regime,
for which the ground state configuration belongs already to the subset $S_z=-2$. 
In addition to several other configurations with $S_z=-1$ we have also performed
calculations for the configuration $1s^2 2s^2 2p_{-1}^2$ belonging to $S_z=0$, 
which is also presented in figure 1. This configuration with the symmetry $^1D$ is the ground state 
of the subset $S_z=0$ at $\gamma=0$. The calculations for this state were necessary in order to
exclude the possibility that it becomes the ground state of the carbon atom 
for some region of the field strength in the weak field case. 

The results of the investigations of this section 
are presented in table \ref{tab:grcon} which contains 
the critical values of $\gamma$ 
at which the crossovers of different ground state configurations take place
(see also figure 5). The critical values of the field strength
given in table I are of course Hartree-Fock values and are expected to shift slightly
for fully correlated calculations.

\section{Selected quantitative aspects}

In table \ref{tab:grenergy} we present the total energies of the seven 
ground state electronic configurations of the carbon atom.
The energy of each state is given at least in the range of field
strength within which it represents the ground state (in many
cases we have performed calculations for a much wider regime of
field strengths). The configurations in table \ref{tab:grenergy} are labeled by 
their corresponding numbers introduced in table \ref{tab:grcon}. 
The data of the tables \ref{tab:grcon} and \ref{tab:grenergy} 
represent the total energy values on a grid of field strengths
covering the complete regime $0\leq\gamma\leq 10000$ and via
interpolation can be used to get the total energies for any field
strength.

The most important work in the literature containing results on the spectrum of the carbon
atom in a strong magnetic field is ref.\cite{JonesOrtiz}. 
We have illustrate the data of ref.\cite{JonesOrtiz} together with our data (see above)
figures 1-4. Unfortunately ref.\cite{JonesOrtiz} does not contain a detailed 
description of the electronic states of the carbon atom considered in their work: 
the data given for electronic states with $S_z=-2$ and $S_z=-3$ 
cannot uniquely be identified with our electronic configurations since the 
quantum numbers of the individual one particle functions of their HF procedure
have not been specified.
Nevertheless we can compare the total energies of the lowest configuration
at a fixed field strength. As a general tendency we observe that in most cases
our energy values are significantly lower than those of ref.\cite{JonesOrtiz}.
Indeed, ref.\cite{JonesOrtiz} obtained only three different ground state
configurations whereas the true picture, as described above, involves
seven different electronic configurations and six associated crossovers.
One can see in figure 1 that it is difficult 
to associate the first three $S_z=-1$ points of ref.\cite{JonesOrtiz} presented 
in this picture with any of our states. 
On the other hand some energy values given in ref.\cite{JonesOrtiz} lie close
to our energies which is the case for the configuration $ 1s^2 2s 2p_0 2p_{-1}^2 $ and we 
perform the comparison with our results for this state in table \ref{tab:Jones1}. 

A second source of data on the carbon atom in strong magnetic fields is
ref.\cite{Neuhauser}. This work deals exclusively with the high field regime
in which the adiabatic approximation represents a reasonable approach
to the wave functions and spectrum. The total energies of the ground state of the
high field limit are given in ref.\cite{Neuhauser} for four different values
of the field strength in
the regime $42.544\le\gamma\le2127.2$. Table IV contains a comparison of those
data with our HF data for the high field ground state configuration
$1s2p_{-1}3d_{-2}4f_{-3}5g_{-4}6h_{-5}$. The third and fourth column represent
our total energies and those of ref.\cite{Neuhauser}, respectively.
As a general tendency our energies are of the order of magnitude of $0.1keV$ lower than those
of ref.\cite{Neuhauser}. This is a nonnegligible deviation in particular
in view of the fact that it exceeds the energy difference to the first excited 
state shown in column five of table IV and for two points of the lowest
field strength the situation is even worse: it exceeds the ionization
energy of the atom, i.e. the energy necessary to build the single positive charged
ion $C^{+}$ (see column six in table IV). We believe that our results are much more accurate than
those of ref.\cite{Neuhauser} since we do not involve any kind of adiabatic
approximation and deal with a flexible wave function due to the flexible mesh calculation
of our one particle functions.

Figure 6 allows us to add some details to the considerations of the previous section. 
This figure presents spatial distributions of the total electronic densities
for the ground state configurations of the carbon atom. 
More exactly, this picture allows us to gain insights into the
geometry of the distribution of the electron density in space and in particular
its dependence on the magnetic quantum number and the total spin.
Thereby we can understand the corresponding impact on the total energy of the atom. 
The first picture in this figure presents the distribution of 
the electron density in the ground state of the carbon atom 
at $\gamma=0$. The following pictures show the distributions of the electronic densities 
at values of the field strength which mark the boundaries of the regimes of field
strengths belonging to the different ground state configurations.
For the high-field ground state we present the distribution of 
the electronic density at the crossover field strength $\gamma=18.664$ 
and for three additional values of $\gamma$ up to $\gamma=1000$. 

For each configuration the effect of the increasing field strength
consists in compressing the electronic distribution towards the $z$ axis.
However most of the crossovers of ground state configurations 
involve the opposite effect which is due to the fact that they  
are associated with an increase of the total magnetic quantum number 
$M={\sum_{\mu=1}^6m_{\mu}}$.

Two issues concerning the results presented above have to be discussed.
First, our HF results do not include the effects of correlation. To take into account the latter
would require a multiconfigurational approach which goes beyond
the scope of the present paper. We, however, do not expect that
the correlation energy changes our main conclusions like, for example, the
fact of the crossovers with respect to the different ground states configurations. 
With increasing field strength the effective one particle
picture should be an increasingly better description of the wave function
and the percentage of the correlation energy should therefore
decrease (see in particular ref.\cite{Schm98b}).
The concrete values of quantities like, for example, the
transition field strengths or the ionization energies, depend however to some
extent on the approach used (Hartree-Fock, correlated methods).

Additional consideration are however required for the crossover situation itself
which separates different electronic configurations of the ground state
and for the other intersections presented in figures 1 to 4.
In principle, effects of electronic correlation can turn
level crossings into avoided crossings.
In this case the classification of the ground state via
a single configuration of six one particle states should
break down and the discontinuous changes of the slope of the total energies
at the points of crossovers should be smoothened out.
But we do not expect these effects of correlation to occur
for the ground state configurations due to the different symmetries of the configurations involved.
Indeed, one can see in table \ref{tab:grcon} that all the neighboring 
ground state configurations have at least either different total magnetic quantum numbers 
or different total spins. Thus, we do not expect that correlation effects
can lead to avoided crossings in the energies of the ground state configurations
of the carbon atom.

The second issue relates to effects of the finite nuclear mass.
For the case of hydrogen it is well-known that in the high field regime ($\gamma >> 10^2$)
mass correction terms due to the finite nuclear mass become relevant
i.e. are no more negligible in
comparison with the Coulomb binding energies. The most important mass
corrections can be included by replacing the electron mass through its reduced
mass and results from the infinite nuclear mass calculations are related
to those with the reduced mass via a scaling relation \cite{Rud94,Bec99}. In the case of the
much heavier C atom these effects are expected to be suppressed significantly due
to the large nuclear mass. In addition the total Coulomb binding energy increases rapidly with
increasing nuclear charge number and is therefore for the carbon atom almost two orders of
magnitude larger compared to the hydrogen atom. This makes the effects of the finite
nuclear mass less important than for hydrogen.

\section{Summary and conclusions}

We have applied our 2D mesh Hartree-Fock method to the magnetized carbon atom.
The method is flexible enough to yield precise results for arbitrary
field strengths and our calculations for the ground and several excited
states are performed for magnetic field strengths ranging from zero
up to  $2.3505\cdot 10^9$T ($\gamma=10000$).
Our considerations focused on the ground state and its crossovers with increasing field strength.
It undergoes six transitions involving seven different electronic configurations.
For weak fields up to $\gamma=0.1862$ the ground state arises 
from the field-free ground state configuration 
$1s^2 2s^2 2p_0 2p_{-1}$ with the total spin $z$-projection $S_z=-1$. 
With increasing strength of the field four different electronic configurations with 
$S_z=-2$ consequently become the ground state: 
$1s^22s2p_0 2p_{-1}2p_{+1}$, $1s^22s2p_0 2p_{-1}3d_{-2}$, 
$1s^22p_0 2p_{-1}3d_{-2}4f_{-3}$ and finally $1s^22p_{-1}3d_{-2}4f_{-3}5g_{-4}$.
At $\gamma=12.216$ we observe the first fully spin polarized configuration
$1s2p_02p_{-1}3d_{-2}4f_{-3}5g_{-4}$ with $S_z=-3$ to become the ground state of the carbon atom. 
At $\gamma=18.664$ the last crossover of the ground state configurations
takes place and ergo for $\gamma>18.664$ the ground state wave function is represented by the 
high-field-limit configuration $1s2p_{-1}3d_{-2}4f_{-3}5g_{-4}6h_{-5}$, $S_z=-3$. 

Our investigation represents the first conclusive study of the ground state
of the carbon atom for arbitrary field strengths. We have obtained a rather
intricate sequence of electronic configurations with increasing field strength.
This underlines the conjecture that the scenario of ground state crossovers with
changing field strength complicates rapidly if we consider neutral atoms with
increasing nuclear charge number. 
Our computations have been performed in the unrestricted Hartree-Fock approximation. 
For the configurations with $S_z=-1$ and $S_z=-2$ (not for those with $S_z=-3$) this means that 
our one determinantal HF wave functions are not eigenfunctions of the operator of 
the total spin. An immediate improvement of our approach would therefore
require a multiconfigurational study. The development of such a code 
capable of describing the wave function in arbitrarily strong magnetic
fields is however a major task and clearly goes beyond
the scope of the present investigation. Putting together what we currently know about
ground states of atoms in strong magnetic fields we can conclude that the H, He, Li and C
atomic ground states have been identified. This leaves plenty of questions open about the
possible ground state configurations of other atoms. 

\vspace*{0.5cm}

\begin{center}
{\bf{Acknowledgments}}
\end{center}
One of the authors (M.V.I.) gratefully acknowledges financial support from the
Deutsche Forschungsgemeinschaft. P.S. acknowledges financial travel support by the
NSF/DAAD and the hospitality of the Department of Physics of the University of Nevada at Reno.

{}

\vspace*{0.5cm}

{\bf Figure Captions}

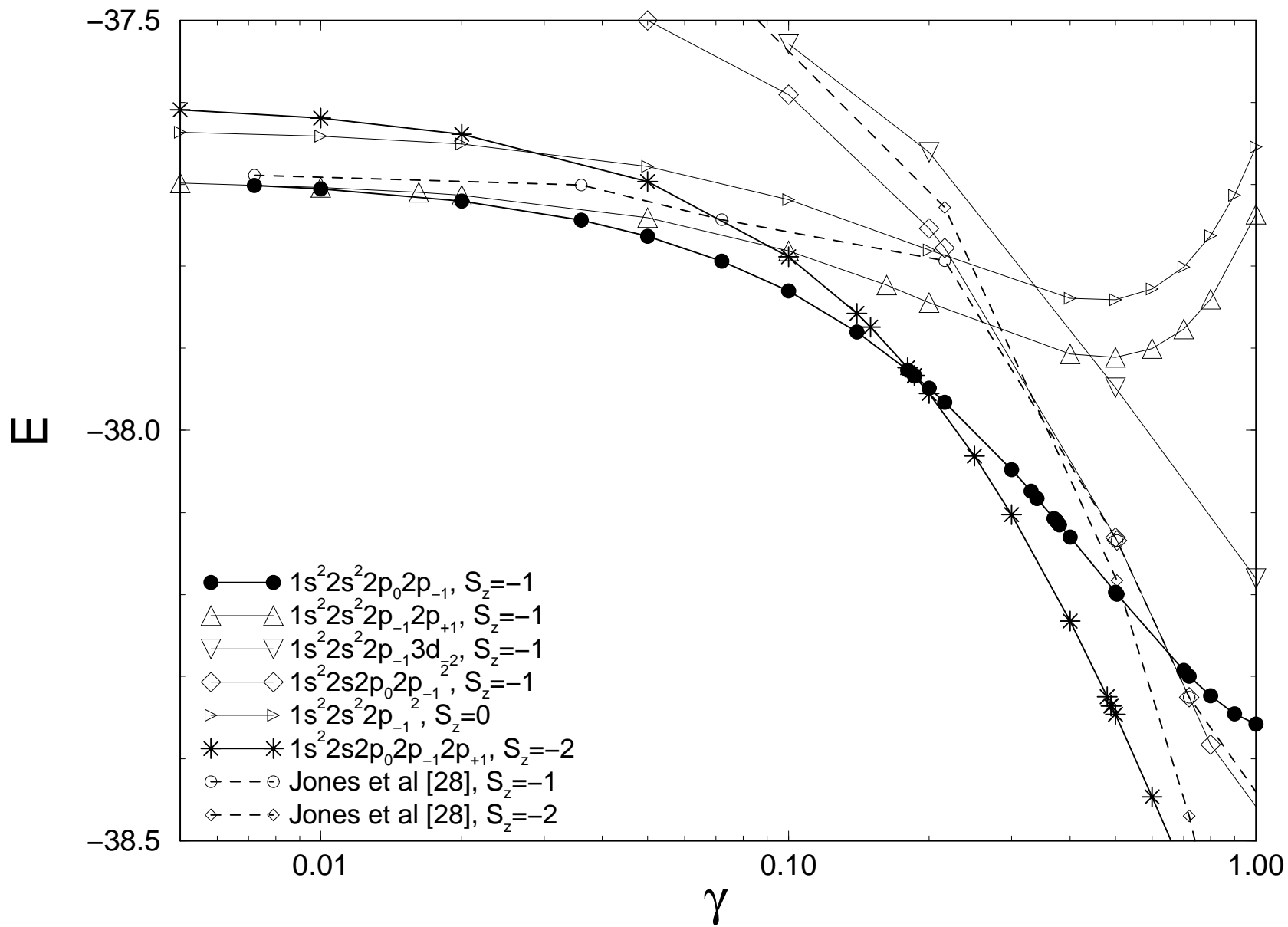
{\bf Figure 1.} The total energies (in atomic units) of the states 
of the carbon atom as functions
of the magnetic field strength 
considered for the determination of the ground state electronic configurations 
with $S_z=-1$. Our results (solid lines) and data taken from ref.\cite{JonesOrtiz} (broken lines).
Energies and field strengths are given in atomic units.

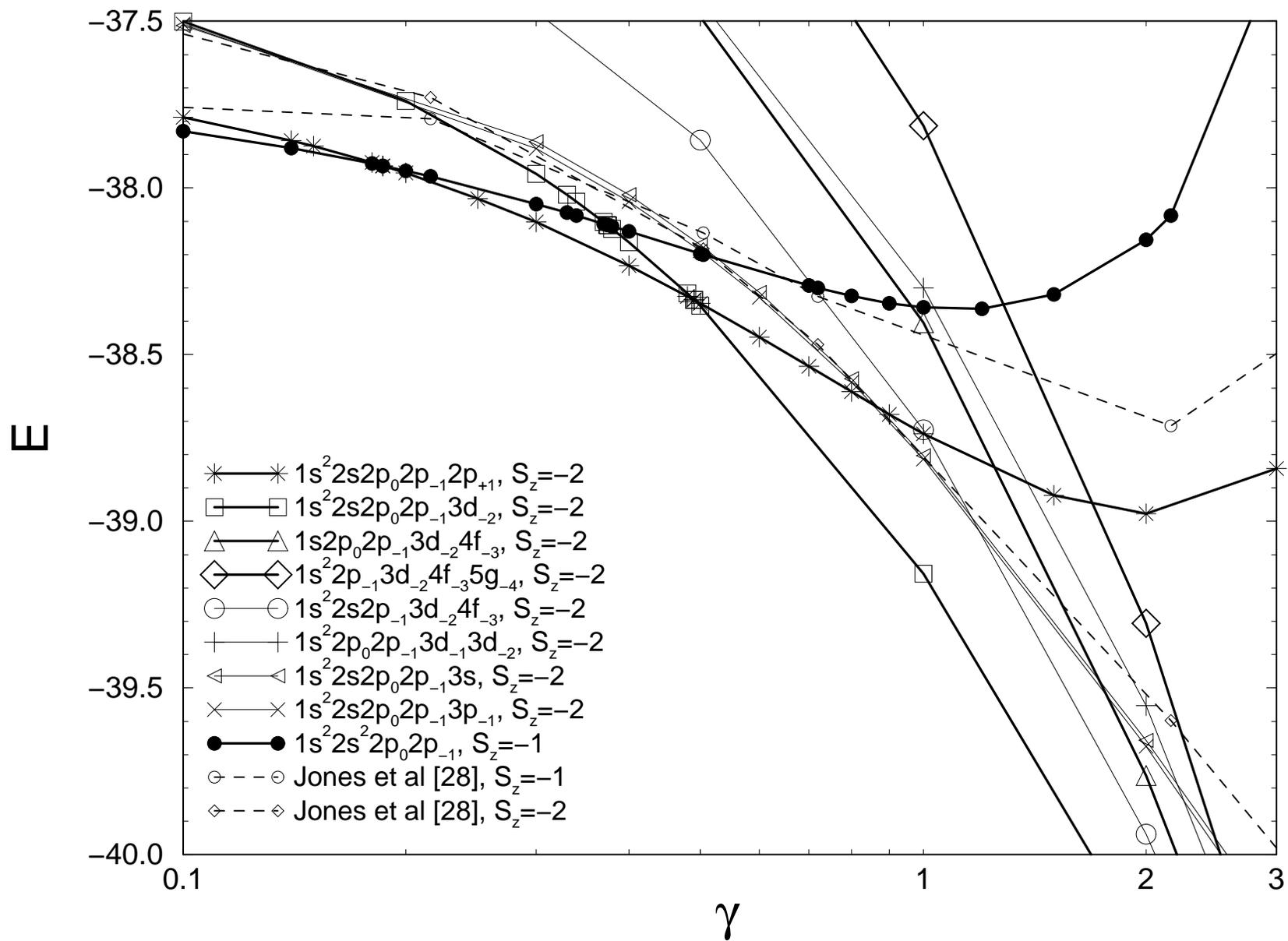
{\bf Figure 2.} Same as in figure 1 for $S_z=-2$, lower field part.
Energies and field strengths are given in atomic units.

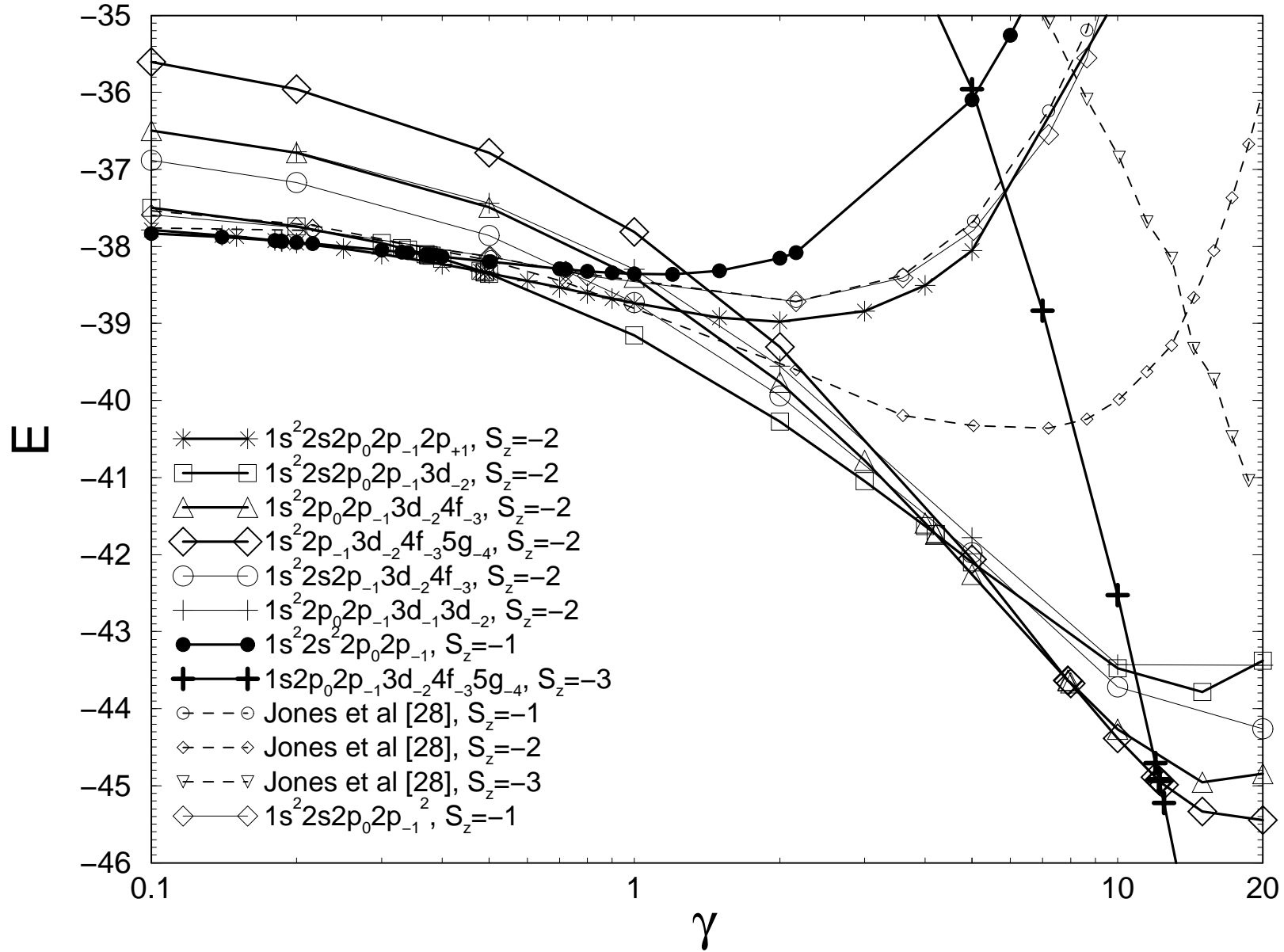
{\bf Figure 3.} Same as in figure 1 for $S_z=-2$, higher field part.
Energies and field strengths are given in atomic units.

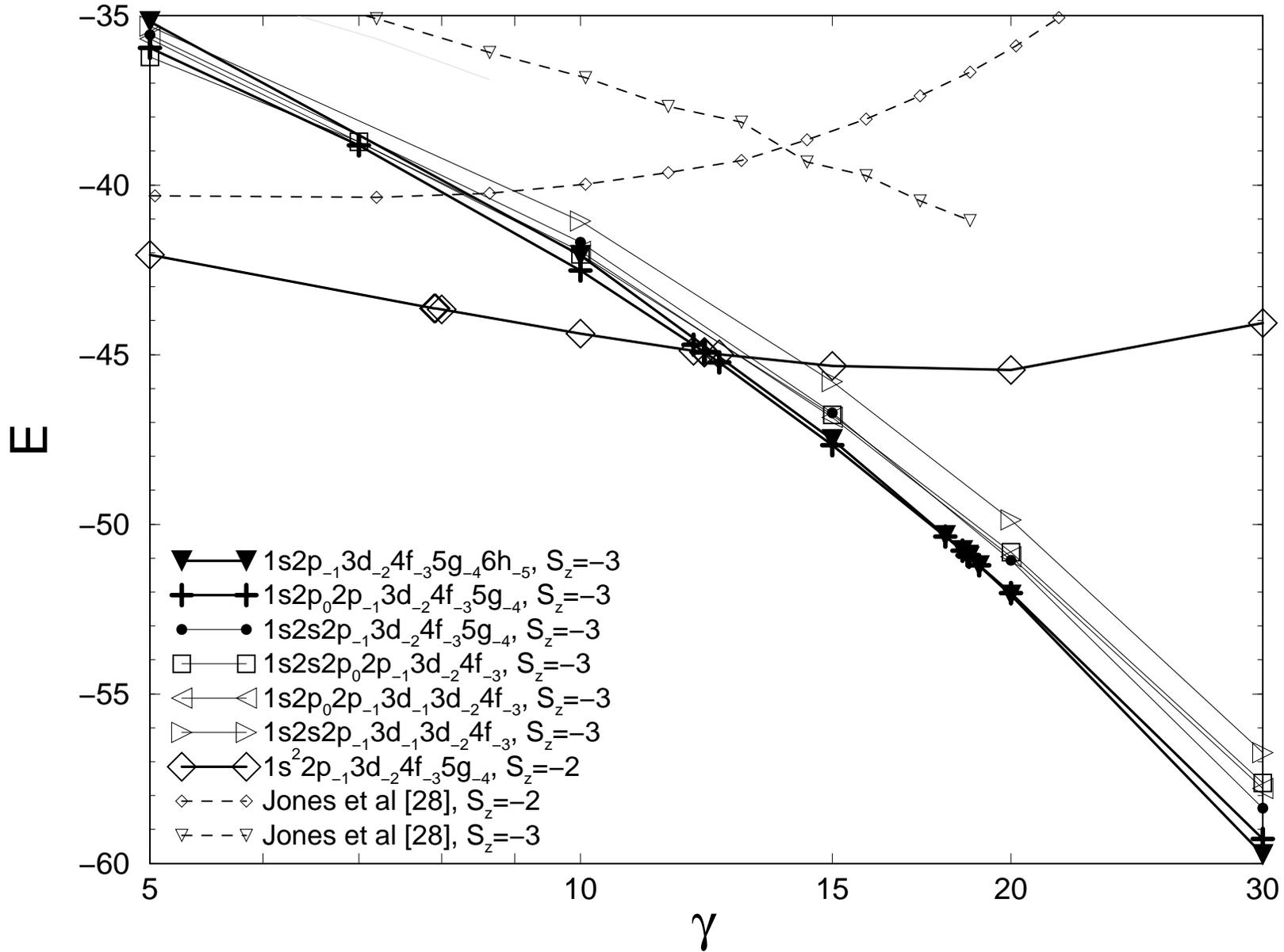
{\bf Figure 4.} Same as in figure 1 for $S_z=-3$.
Energies and field strengths are given in atomic units.

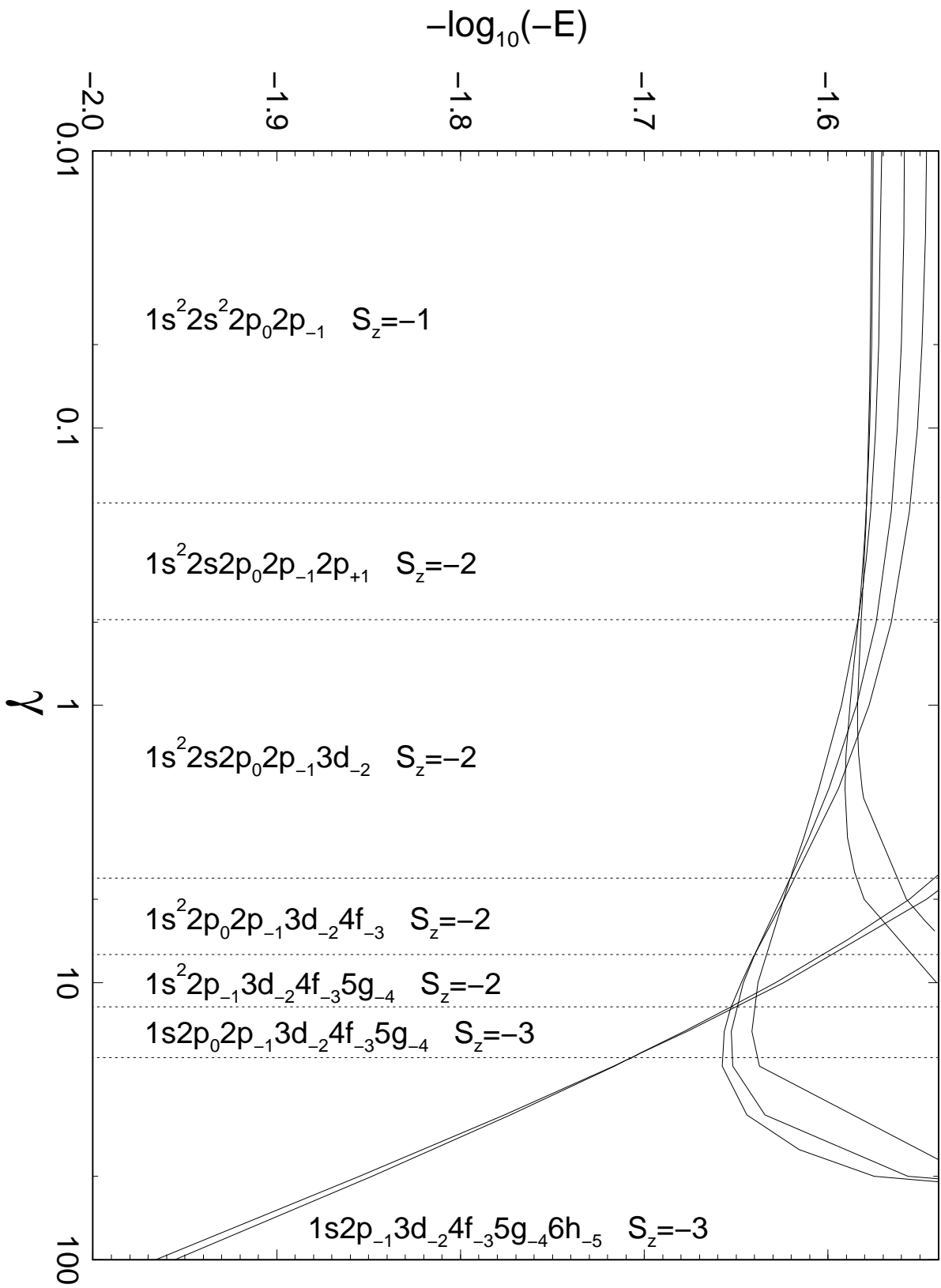
{\bf Figure 5.} Energies of the ground state configurations
as a function of the field strength. Vertical dotted lines
divide regions belonging to different Hartree-Fock 
ground state configurations. 
Energies and field strengths are given in atomic units.

{\bf Figure 6.} Contour plots of the total electronic
densities for the ground state of the carbon atom.
For neighboring lines the densities are different by a factor of $e$.
The coordinates $z$, $\rho$ as well as the
corresponding field strengths are given in atomic units.

\newpage
\begin{table}
\caption{The Hartree-Fock ground state configurations of the carbon atom 
in external magnetic fields. 
The configurations, presented in the table are the ground state configurations 
at $\gamma_{\rm min}\leq\gamma\leq\gamma_{\rm max}$.}
\begin{tabular}{@{}llllrllllll}
no.&$\gamma_{\rm min}$&$\gamma_{\rm max}$& The ground state configuration&$M$&$S_z$&$E(\gamma_{\rm min})$\\
\noalign{\hrule}
1&0     &0.1862   &$1s^2 2s^2 2p_0 2p_{-1}$                 &$-1$ &$-1$&$-37.69096$\\
2&0.1862&0.4903   &$1s^22s2p_0 2p_{-1}2p_{+1}$              &$0 $ &$-2$&$-37.9334$\\
3&0.4903&4.207    &$1s^22s2p_0 2p_{-1}3d_{-2}$              &$-3$ &$-2$&$-38.3359$\\
4&4.207 &7.920    &$1s^22p_0 2p_{-1}3d_{-2}4f_{-3}$         &$-6$ &$-2$&$-41.7369$\\
5&7.920 &12.216   &$1s^22p_{-1}3d_{-2}4f_{-3}5g_{-4}$       &$-10$&$-2$&$-43.6397$\\
6&12.216&18.664   &$1s2p_02p_{-1}3d_{-2}4f_{-3}5g_{-4}$     &$-10$&$-3$&$-44.9341$\\
7&18.664&$\infty$ &$1s2p_{-1}3d_{-2}4f_{-3}5g_{-4}6h_{-5}$  &$-15$&$-3$&$-50.9257$\\
\end{tabular}
\label{tab:grcon}
\end{table}

\newpage
\begin{table}
\caption{The energies of the ground state configurations of the carbon atom 
dependent on the magnetic field strength. 
The figures in paratheses are the numbers of the ground state configurations 
provided in the first column of table I.
Energies and field strengths are given in atomic units.}
\begin{tabular}{@{}llllllllll}
$\gamma$&$E(1)   $&$E(2)     $&$E(3)    $&$E(4)    $&$E(5)    $&$E(6)    $&$E(7)    $\\
\noalign{\hrule}
0.000 &$-37.69096$&$-37.59928$&$-37.2188$&$-36.1170$&$-35.1186$&$-24.3112$&$-22.5588$\\
0.001 &$-37.6925$&$-37.6013$&$  -37.2224$&$-36.122 $&$-35.126 $&$-24.3193$&$-22.570 $\\
0.002 &$-37.6940$&$-37.6034$&$  -37.2259$&$-36.1270$&$-35.133 $&$-24.3272$&$-22.580 $\\
0.005 &$-37.6985$&$-37.6094$&$  -37.2360$&$-36.1413$&$-35.152 $&$-24.3496$&$-22.608 $\\
0.01  &$-37.7059$&$-37.6193$&$  -37.2526$&$-36.1641$&$-35.1827$&$-24.3853$&$-22.6539$\\  
0.02  &$-37.7205$&$-37.6389$&$  -37.2842$&$-36.2085$&$-35.2399$&$-24.4525$&$-22.7376$\\
0.05  &$-37.7633$&$-37.6966$&$  -37.3705$&$-36.3247$&$-35.3900$&$-24.6328$&$-22.9579$\\
0.1   &$-37.8302$&$-37.7882$&$  -37.5010$&$-36.4932$&$-35.6027$&$-24.8959$&$-23.2743$\\ 
0.2   &$-37.9486$&$-37.9552$&$  -37.7388$&$-36.7835$&$-35.9600$&$-25.3546$&$-23.8125$\\ 
0.3   &$-38.0479$&$-38.1026$&$  -37.9579$&$-37.0387$&$-36.2653$&$-25.7617$&$-24.2813$\\
0.4   &$-38.1302$&$-38.2323$&$  -38.1624$&$-37.2716$&$-36.5376$&$-26.1358$&$-24.7067$\\
0.5   &$-38.1973$&$-38.3464$&$  -38.3541$&$-37.4881$&$-36.7864$&$-26.4865$&$-25.1007$\\
0.6   &$-38.2510$&$-38.4467$&$  -38.5339$&$-37.6912$&$-37.0170$&$-26.8187$&$-25.4706$\\
0.7   &$-38.2927$&$-38.5346$&$  -38.7033$&$-37.8830$&$-37.2329$&$-27.1362$&$-25.8211$\\
0.8   &$-38.3238$&$-38.6116$&$  -38.8632$&$-38.0650$&$-37.4364$&$-27.4411$&$-26.1554$\\
0.9   &$-38.3453$&$-38.6788$&$  -39.0145$&$-38.2385$&$-37.6293$&$-27.7351$&$-26.4758$\\ 
1.    &$-38.3582$&$-38.7373$&$  -39.1577$&$-38.4043$&$-37.8130$&$-28.0195$&$-26.7843$\\ 
1.5   &$-38.3192$&$-38.9225$&$  -39.7776$&$-39.1406$&$-38.6242$&$-29.3275$&$-28.1888$\\  
2.    &$-38.1549$&$-38.9770$&$  -40.2769$&$-39.7621$&$-39.3061$&$-30.4938$&$-29.4282$\\
3.    &$-37.6035$&$-38.8418$&$  -41.0477$&$-40.7780$&$-40.4222$&$-32.5445$&$-31.5925$\\
4.    &$-36.8901$&$-38.5088$&$  -41.6319$&$-41.5886$&$-41.3180$&$-34.3404$&$-33.4807$\\
5.    &$-36.0969$&$-38.0551$&$  -42.1016$&$-42.2549$&$-42.0608$&$-35.9601$&$-35.1815$\\
7.    &$        $&$-36.9093$&$  -42.8151$&$-43.2771$&$-43.2195$&$-38.8325$&$-38.1971$\\
8.    &$        $&$-36.2426$&$  -43.0841$&$-43.6685$&$-43.6734$&$-40.1320$&$-39.5620$\\
10.   &$        $&$-34.7539$&$  -43.4769$&$-44.2659$&$-44.3872$&$-42.5277$&$-42.0799$\\
12.   &$        $&$        $&$  -43.7032$&$-44.6615$&$-44.8905$&$-44.7094$&$-44.3749$\\
15.   &$        $&$        $&$  -43.7842$&$-44.9571$&$-45.3348$&$-47.6770$&$-47.5002$\\
20.   &$        $&$        $&$  -43.3767$&$-44.8468$&$-45.4465$&$-52.0282$&$-52.0890$\\
30.   &$        $&$        $&$  -41.1418$&$-43.0912$&$-44.0660$&$-59.2747$&$-59.7433$\\
40.   &$        $&$        $&$  -37.6519$&$-39.9996$&$-41.2855$&$-65.2949$&$-66.1073$\\
50.   &$        $&$        $&$  -33.3211$&$-36.0197$&$-37.5718$&$-70.5187$&$-71.6285$\\
100.  &$        $&$        $&$          $&$\ -8.865$&$-11.3693$&$-90.2751$&$-92.4552$\\
200.  &$        $&$        $&$          $&$+60.973 $&$+57.3384$&$-116.4070$&$-119.8127$\\
500.  &$        $&$        $&$          $&$+307.31 $&$+301.1777$&$-163.209$&$-168.5248$\\
1000. &$        $&$        $&$          $&$+754.1  $&$+746.589$&$-209.98  $&$-217.1413$\\
2000. &$        $&$        $&$          $&$        $&$        $&$-268.711 $&$-278.1612$\\
5000. &$        $&$        $&$          $&$        $&$        $&$-368.1   $&$-381.8097$\\
10000.&$        $&$        $&$          $&$        $&$        $&$-463.7   $&$-480.875$\\
\end{tabular}
\label{tab:grenergy}
\end{table}

\begin{table}
\caption{The energy of the state $1s^2 2s 2p_0 2p_{-1}^2$ of the carbon atom 
in magnetic fields compared with results by Jones et al(1996). Energies and
field strengths are given in atomic units.}
\begin{tabular}{@{}llllllllll}
$\gamma$&$E(1s^2 2s 2p_0 2p_{-1}^2)$&$E$\cite{JonesOrtiz} $S_z=-1$\\
\noalign{\hrule}
0.216 &$-37.7779$   &$-37.793$\\
0.504 &$-38.1345$   &$-38.134$\\
0.720 &$-38.3254$   &$-38.325$\\
2.16  &$-38.7188$   &$-38.713$\\
3.6   &$-38.4076$   &$-38.376$\\
5.04  &$-37.7940$   &$-37.677$\\
7.2   &$-36.5467$   &$-36.239$\\
8.64  &$-35.5527$   &$-35.196$\\ 
14.4  &$-30.6343$   &$-29.969$\\
21.6  &$-23.0362$   &$-21.472$\\
\end{tabular}
\label{tab:Jones1}
\end{table}

\begin{table}
\caption{The energies of the high-field ground state of the carbon atom 
$E({\rm C})$, its first high-field 
excited state ($1s2p_{-1}3d_{-2}4f_{-3}5g_{-4}7i_{-6}$)
and the energy of the ion ${\rm C^+}$ compared
with the adiabatic results by Neuhauser et al (1987) for 
the high-field ground state.}
\begin{tabular}{@{}llllllllll}
$B\ (10^{12}$G)&$\gamma$&$E({\rm C})$ (keV)&$E({\rm C})$\cite{Neuhauser} (keV)
&$E(1s2p_{-1}3d_{-2}4f_{-3}5g_{-4}7i_{-6})$ (keV)
&$E({\rm C^+})$ (keV)
\\
\noalign{\hrule}
0.1 &42.544&$-1.83895 $   &$-1.678$&$-1.82464$&$-1.78348$\\
0.5 &212.72&$-3.33639 $   &$-3.22 $&$-3.30957$&$-3.22647$\\
1.0 &425.44&$-4.31991 $   &$-4.20 $&$-4.28530$&$-4.17396$\\
5.0 &2127.2&$-7.73528 $   &$-7.60 $&$-7.67499$&$-7.46051$\\
\end{tabular}
\label{tab:Neuhtab}
\end{table}

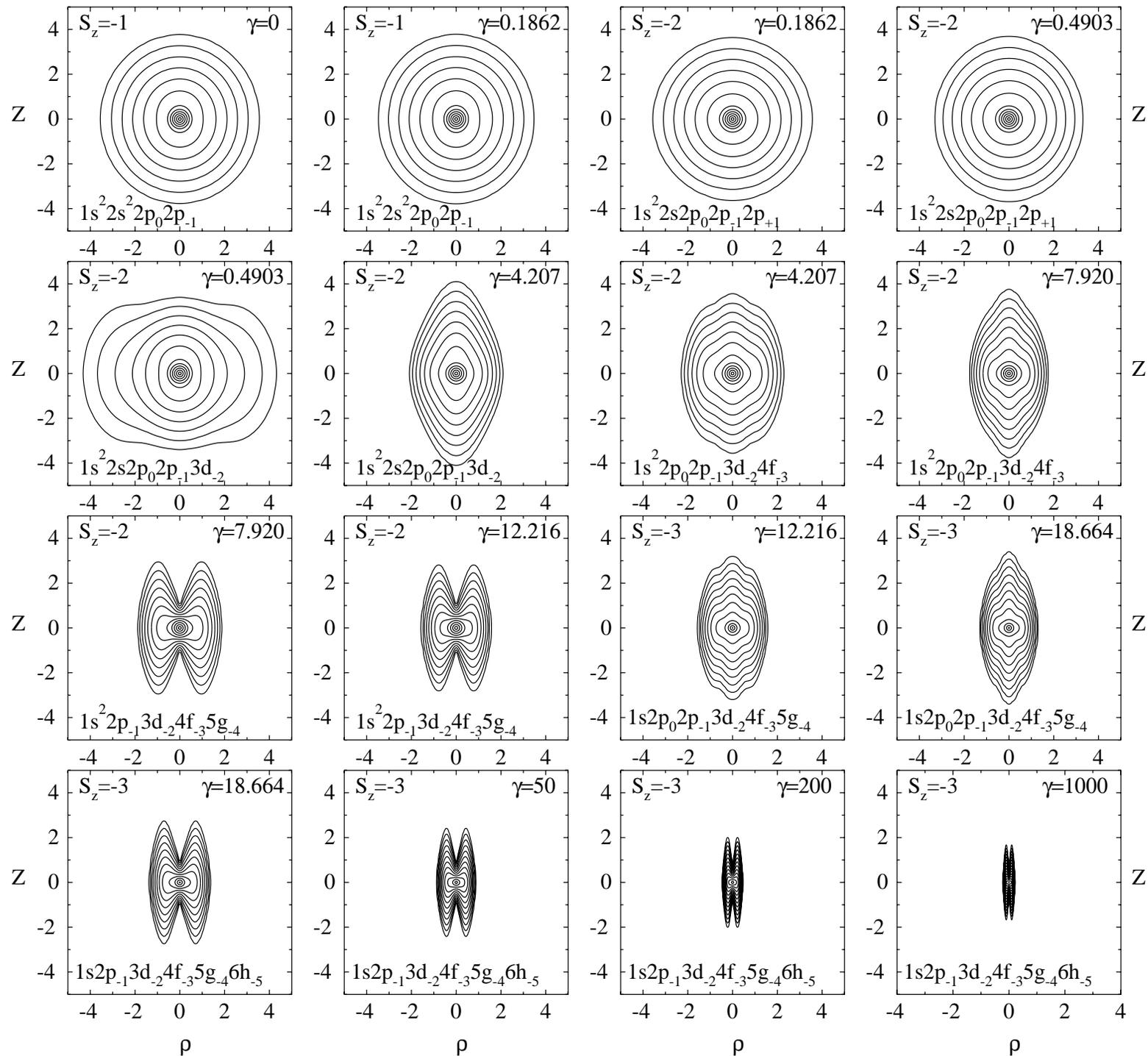
\end{document}